\documentclass[12pt]{iopart}
\usepackage{amssymb,amsthm}

\begin{document}

\title{Conformally invariant cosmology based on Riemann--Cartan spacetime}

\author{Yuri V Shtanov$^{a,b}$ and Sergei A Yushchenko$^a$}

\address{$^a$Bogolyubov Institute for Theoretical Physics, Kiev 03680,
Ukraine}
\address{$^b$Department of  Physics, Brown University, Providence, RI 02912, USA}
\eads{\mailto{shtanov@bitp.kiev.ua}, \mailto{yushchenko@bitp.kiev.ua}}

\date{18 July 1994}

\begin{abstract}
\noindent Conformally invariant GUT-like model including  gravity based on
Riemann--Cartan  space-time  $U_4$ is considered. Cosmological scenario that follows from
the model is discussed and standard quantum  gravitational  formalism in the
Arnowitt--Deser--Misner form is developed. General formalism is then illustrated on
Bianchi-IX minisuperspace cosmological  model. Wave functions  of  the universe in the de
Sitter minisuperspace model with Vilenkin and Hartle--Hawking boundary conditions are
considered and corresponding  probability  distributions  for the scalar field values are
calculated.
\end{abstract}

\pacs{04.60-m, 98.80.Qc \bigskip \\
{\em Preprint\/} BROWN-HET-937}

\maketitle

\section{Introduction}

In unifying gravity  with  other  interactions  concepts  of
symmetries, undoubtedly, play an important role. Among various
symmetries that
have been put  forward  at  different  times  the  symmetry  with
respect  to  local  conformal  transformations remains to be of
special significance and keeps  attracting
attention of many researches. First, because theories with such a
symmetry typically exhibit better quantum behaviour, and the issue
of renormalizability might lead to conformal invariance. Second,
because such kind of symmetry might help to resolve the cosmological
constant problem (the usual $\Lambda$-term breaks the conformal symmetry
of the action and therefore is excluded simply by a demand of such
a symmetry).

The original idea of local conformal invariance belongs to  Weyl [1] who introduced a
compensating  gauge  vector  field  for  the transformations under consideration. The
corresponding  geometric structure is usually called Weylian. It was  afterwards
developed in various directions [2--10].

Invariance with respect to local  conformal  transformations can  be  implemented  into
theory  of  gravity  in different manners [11]. One of the rather simple ways  is  to
base  a theory on Riemann-Cartan spacetime which naturally arises  within the framework
of the Poincar\'e gauge theory of gravity. Some  work has already been done along these
lines.  Conformally  invariant dynamics of spinor fields on a  background  Riemann-Cartan
space were studied in [12]. In the works [13] a simple  example  of  the gravitational
part  of  a  theory  of  such  a  kind  has   been considered. It incorporated a scalar
field into the term $\varphi^2 R$ in the Lagrangian. The aim of the present  paper  is
to  generalize these  proposals  and  to  consider  a  realistic  theory   which
possesses Weyl invariance. In doing this  we  take  into  account that the vector trace
$Q_\mu$  of the torsion tensor  transforms  under the action of the local conformal group
similarly  to  the  Weyl vector. Making use  of  this  property  we  construct  a
generic GUT-like model, based on the Riemann--Cartan geometric  structure, which contains
a scalar field multiplet with its kinetic term  in the Lagrangian and which  is
invariant  with  respect  to  local conformal transformations.  The  key  difference  of
the  theory considered here from the theory of Weyl is that in the last one a special
vector field is introduced as a compensating gauge  field for the local conformal
transformations  whereas  in  our  theory this role is played by the torsion  trace
vector.  We  therefore avoid to introduce new specific entities to make  our  theory
conformally invariant.

The role of the scalar fields in the theory considered here,
on the one hand, is the same as in ordinary GUT models. They give
masses to  gauge  vector  bosons  and  to  fermions  through  the
symmetry breaking mechanism  of  Higgs,  making  it  possible  to
preserve the (ordinary) gauge invariance of the  theory.  On  the
other hand, they allow one to use the term of  type $\varphi^2 R$
in  the
Lagrangian  and  thus  to  generate  the  gravitational  coupling
constant. In such a way the theory is extended to include gravity
and,  at  the  same  time,  remains  to  be  locally  conformally
invariant. In this paper we  will  consider  some  of  the  basic
features of such  a  theory,  in  particular,  those  related  to
inflationary and quantum cosmology.

We will see below that the cosmological constant problem is not
actually solved in our theory. However, the value of $\Lambda$
in the effective $\Lambda$-term
is now not an independent parameter, but is a function of some other
parameters of the model. It is therefore at least restricted and
further investigation might discover the possibility  of having it equal
to zero without fine tuning the constants of the theory.

The paper is organized as follows. In the  next  section  we present a locally
conformally  invariant  version  of  a  generic GUT-like model coupled to gravity. In
Section  3  a  cosmological inflationary scenario  based  on  our  theory  is  discussed.
In Section 4 we develop a standard quantum cosmological formalism in the
Arnowitt--Deser--Misner (ADM) form for the  theory  considered. General formalism will
then be illustrated  in  Section  5  on  a minisuperspace cosmological model  of
Bianchi-IX  type.  In  the Appendix we provide the necessary geometric background
including the description of the  Riemann--Cartan  geometric  structure  and local
conformal transformations.

\section{Conformally invariant theory}

Our theory will be based on Riemann--Cartan space $U_4$  (for more detailed description
and basic notations see the Appendix). Riemann--Cartan structure implies the presence of
the affine connection form $\omega^a{}_b$ and the metric tensor $g$ which is covariantly
constant:
$$
\nabla^\omega g  =  0 \, , \eqno (1)
$$
where the symbol $\nabla^\omega$ denotes the covariant derivative
specified  by
the affine connection $\omega^a{}_b$. The affine connection is  supposed  to
satisfy only the metricity condition (1) hence torsion tensor  is
not assumed to vanish.

We are going to consider  a  rather  natural  conformally  invariant
generalization of GUT-like model with the Lagrangian of the form
$$
L  = {1\over 2} m^2 (\varphi) R (\omega) + {i\over 2} \left( \overline
\psi \gamma^{\mu}{{\cal D}}_{\mu} \psi - \overline{({\cal D}_{\mu} \psi)}
\gamma^{\mu} \psi \right) - \{ f \varphi \overline \psi \psi \}
$$
$$
{} - {1\over 2} |{\cal D} \varphi|^2  - V(\varphi) - {1\over 4 e^2}
Tr F_{\mu\nu} F^{\mu\nu} \, . \eqno                 (2)
$$
It is  constructed  from  a  multiplet $\psi$  of  spinor  fields,  a multiplet $\varphi$
of scalar (Higgs) fields (multiplet indices of $\varphi$  and $\psi$ will often be
omitted), gauge connection field  $A  =  A_\mu dx^\mu$ of some gauge group $G$, metric $g
= g_{\mu\nu} dx^\mu \otimes dx^\nu = \eta_{ab} e^a \otimes e^b$  which can be presented
by the orthonormal tetrad  components $e^a_\mu$ or $e^\mu_a$, and affine connection field
$\omega^a{}_b = \omega^a{}_{b\mu} dx^\mu = \omega^a{}_{bc} e^c$.  The  metric signature
is taken to be $(-,+,+,+)$. Greek indices are raised  and lowered by the metric
components $g_{\mu\nu}$ and $g^{\mu\nu}$   in  the  coordinate base, and Latin indices -
by the metric components $\eta_{ab}$ and $\eta^{ab}$   in the orthonormal tetrad base.
$R(\omega)$ is the Riemann--Cartan curvature scalar. "Long" derivatives ${\cal D}_\mu$
which enter the Lagrangian (2)  are defined by the equations
$$
{\cal D}_\mu \psi = (\partial_\mu + \omega_\mu - Q_\mu + A_\mu )
\psi \, , \eqno  (3)
$$
$$
{\cal D}_\mu \varphi = (\partial_\mu - {2\over 3} Q_\mu + A_\mu )
\varphi \, , \eqno     (4)
$$
where $Q_\mu$  is the vector trace of the torsion tensor:  $Q_\mu
=  Q^\alpha{}_{\mu\alpha}$,
and
$$
\omega = - {1\over 8} \omega^{ab}\, [\gamma_a , \gamma_b ] \eqno  (5)
$$
is the spin connection form. $\gamma^a$ are  the  usual  constant  Dirac
matrices. The quantities $F_{\mu\nu}$ are the components of the  curvature
two-form of the gauge connection form $A = A_\mu  dx^\mu$:
$$
iF =  dA + A \wedge A \, . \eqno                       (6)
$$
The symbol $\{f\varphi \overline \psi \psi\}$ denotes the  sum  of
various  possible  Yukawa
couplings between the spinor and the scalar Higgs multiplets with
coupling constants $\{f\}$. The values $m^2 (\varphi)$ and $V(\varphi)$
are assumed  to
be analytic in $\varphi$, so in order to preserve conformal invariance of
the action (about the conformal transformations see  below)  they
must represent respectively a quadratic and a quartic forms of $\varphi$.
We  assume  $m^2 (\varphi)$  to  be  positive  definite.
Thus for the multiplet $\{\varphi^A$, $A = 1,\ldots,k\}$  of
real scalar fields we put
$$
m^2 (\varphi) = \xi_{AB}\, \varphi^A \varphi^B \, , \eqno    (7)
$$
where $m^2 (\varphi) > 0$  for $\varphi \not= 0$, and
$$
V(\varphi) = \lambda_{ABCD}\, \varphi^A \varphi^B \varphi^C \varphi^D
\, , \eqno                        (8)
$$
where $\xi_{AB}$ and $\lambda_{ABCD}$ are real dimensionless
constants symmetric in
their indices. The functions $m^2 (\varphi)$, $V(\varphi)$ and
$\{f\varphi\overline \psi\psi\}$  are  assumed
also to be $G$-invariant. Finally, the trace in (2) is taken in the
representation space of the group $G$.

The action of the theory is written as
$$
S  = \int\limits_M L \sqrt{-g} d^4 x + \int\limits_{\partial M}
m^2 (\varphi) {\cal K}(\omega) \sqrt{h} d^3 x \, , \eqno  (9)
$$
where ${\cal K}(\omega)$ is the scalar extrinsic curvature
(with respect to the
connection $\omega^a{}_b$) of the boundary $\partial M$ of the integration
region  $M$,
$h_{ij}$ $(i,j = 1,2,3)$ are  the  induced  metric  components  on
this boundary, and $h = det(h_{ij})$.

The action (9) with the Lagrangian (2) was constructed so as to differ as
little as possible from the general relativity action. The only basic
difference between them is the presence of the function $m^2(\varphi)$
instead of a constant $(8\pi G_{\rm N})^{-1}$, and the presence of
torsion in the scalar curvature $R$ and in the metric sector of the theory.

The action (9) is invariant with respect to the  group  of  local  conformal
transformations
$$
g(x) \rightarrow g^\prime (x) = \exp \left( 2 \sigma (x) \right) g(x) \, ,
\eqno (10)
$$
$$
\varphi (x) \rightarrow  \varphi^\prime (x) = \exp \left( - \sigma (x) \right)
\varphi (x) \, , \eqno                      (11)
$$
$$
\psi (x) \rightarrow \psi^\prime (x) = \exp \left( - {3\over 2} \sigma
(x) \right) \psi (x) \, , \eqno                     (12)
$$
$$
\overline \psi (x) \rightarrow \overline \psi^\prime (x) = \exp \left(
- {3\over 2} \sigma (x) \right) \overline \psi (x) \, , \eqno  (13)
$$
$$
A(x) \rightarrow  A^\prime (x) = A(x) \, , \eqno           (14)
$$
$$
\omega^a{}_b(x) \rightarrow \omega^{\prime a}{}_b (x) = \omega^a{}_b (x) \, ,
\eqno (15)
$$
where $\sigma (x)$ is an arbitrary real function. Such an  invariance  is provided in
particular by using  the  ``long"  derivatives  ${\cal D}_\mu$   as defined in (3) and
(4). They  involve  torsion  trace  vector  $Q_\mu$ which under the action of the local
conformal  group  transforms similarly to the Weyl gauge vector field (see the Appendix):
$$
Q_\mu (x) \rightarrow Q^\prime_\mu (x) = Q_\mu (x) - {3\over 2} \partial_\mu
\sigma (x) \, . \eqno                (16)
$$
Due to this property ``long" derivatives  (3)  and  (4)  transform under the local
conformal  group  just  like  the  corresponding (spinor and scalar) fields themselves.
Note that the torsion trace $Q_\mu$ in the derivative (3) of a spinor in fact drops out
of the spinor kinetic term of the Lagrangian (2) due to Hermitian form of the latter.
Conformal  invariance  of the action (9) is also provided by the specific shape (7) and
(8) of the functions $m^2 (\varphi)$ and $V(\varphi)$ respectively.

Both terms of the action (9) are invariant with  respect  to
the local  conformal  transformations  written  just  above.  The
second, boundary, term does not affect the equations  of  motion.
We have added it to the  action  in  order  to  recover,  in  the
natural gauge $m^2 (\varphi) = const$, the usual  Gibbons-Hawking
boundary term [14] of the Hilbert-Einstein action for gravity, as
we  will see a bit later.

The key difference of the theory considered  here  from  the
theory of Weyl is that in the last one a special vector field  is
introduced as a compensating gauge field for the local  conformal
transformations whereas in our theory this role is played by  the
torsion  trace  vector.  We  therefore  use  only  the  geometric
structure already at our disposal  and  avoid  to  introduce  new
entities to make our theory conformally invariant.

One can think of the action (9) with the Lagrangian (2) as of the first lowest terms in
the expansion in field derivatives of some generic conformally invariant action based on
Riemann--Cartan geometry.

The equations of motion are obtained by varying  the  action
(9) over the independent variables $\varphi,\, \psi,\, \overline
\psi,\, A_\mu,\, \omega^a{}_{b\mu},$  and $e^a_\mu$.
Varying over the affine connection components $\omega^a{}_{b\mu}$
we obtain  the
following equations for the torsion tensor
$$
m^2 (\varphi) \left( Q^\mu{}_{\nu\sigma} + {2\over 3} \delta^\mu{}_{[\nu}
Q_{\sigma ]}\right) = {1\over 4} \varepsilon^\mu{}_{\nu\sigma\tau}
\overline \psi \gamma^\tau \gamma^5 \psi \, , \eqno       (17)
$$
$$
Q_\mu = {3\over 4} \partial_\mu \ln M^2 (\varphi) \, , \eqno (18)
$$
where $\varepsilon_{\mu\nu\sigma\tau}$ is  the  antisymmetric  tensor with  the
components $\varepsilon_{\rm 0123} = - \sqrt {- g}$, $\gamma^\mu = e^\mu_a \gamma^a$,
$\gamma^5 =  {i\over 4}\, \varepsilon_{abcd} \,\gamma^a \gamma^b \gamma^c \gamma^d = -
\,i\, \gamma^0 \gamma^1 \gamma^2 \gamma^3$, and
$$
M^2 (\varphi) = m^2 (\varphi) + {1\over 6} \varphi^2 ,\,\,\,
\varphi^2 = \sum_{A=1}^k \varphi^A \varphi^A \, . \eqno   (19)
$$
Purely algebraic equations (17),  (18)  for  the  torsion  tensor
components reflect  the  fact  that  torsion  in  our  theory  is
non-propagating.

At this point we can  substitute  the  expressions  for  the
torsion from (17) and (18) back into the action (9). We will then
obtain the following torsion-free effective action
$$
S\,_{\rm eff} = \int\limits_M L\,_{\rm eff} \sqrt {- g}\, d^4x + \int\limits_
{\partial M} m^2 (\varphi) {\cal K} (\Gamma) \sqrt{h}\, d^3x \, , \eqno (20)
$$
the Lagrangian of which is
$$
L\,_{\rm eff} = {1\over 2} m^2(\varphi) R(\Gamma) + {i\over 2} \left(
\overline \psi \gamma^\mu D_\mu \psi - \overline {(D_\mu \psi)}
\gamma^\mu \psi \right) - \{f \varphi \overline\psi \psi \}
$$
$$
{} - {3\over 16m^2(\varphi)} \left( \overline \psi \gamma^\mu \gamma^5
\psi \right) \left( \overline \psi \gamma_\mu \gamma^5 \psi \right)
- {1\over 2} |D \varphi|^2 + 3 \left( \nabla M(\varphi) \right)^2
- V(\varphi) - {1\over 4e^2} Tr F_{\mu\nu}F^{\mu\nu} \, , \eqno (21)
$$
and in particular contains the term that describes four-fermionic
axial current $\,\times\,$ current  interaction.  Here  $R(\Gamma)$
is  the   usual
Riemannian curvature scalar, and ${\cal K}(\Gamma)$ is the extrinsic
curvature
of the boundary $\partial M$ with respect to the Riemannian connection form
$\Gamma^a{}_b$. The "long" derivatives $D_\mu \psi$ and $D_\mu \varphi$
in (21)  are  defined  as follows
$$
D_\mu \psi = (\partial_\mu + \Gamma_\mu + A_\mu) \psi \, , \eqno (22)
$$
$$
D_\mu \varphi = (\partial_\mu + A_\mu) \varphi \, , \eqno (23)
$$
where
$$
\Gamma_\mu = - {1\over 8} \Gamma^{ab}{}_\mu [\gamma^a,\gamma^b] \eqno (24)
$$
is  the  spin  connection   constructed   from   the   Riemannian
(torsion-free) affine connection $\Gamma^a{}_b$.
The symbol  $\nabla$ will denote the
usual covariant derivative with respect to the affine  connection
$\Gamma^a{}_b$.

The action (20) remains to be locally conformally invariant,
although its two terms are not invariant separately, contrary  to
the expression (9) for the former action. Note that in the  gauge
$m^2 (\varphi) = const$ the second, boundary, term in (20)  reproduces  the
Gibbons-Hawking  boundary  term  [14]  of  the   Hilbert-Einstein
theory.

Equations of motion for the scalar multiplet fields  $\varphi$  that
stem from the action (20)  can  be  put  in  the  following  form
(multiplet indices are  omitted, so, for  example,  the  equality
$\varphi = 0$ denotes that all the multiplet components $\varphi^A$
are zero,  and
$\varphi \not= 0$ means that some of the components are nonzero)
$$
D^\dagger _\mu D^\mu \varphi - {{\nabla^2 M(\varphi)}\over {M(\varphi)}}
\varphi
- {{\partial V(\varphi)}\over {\partial \varphi}} - \{f \overline \psi
\psi \}
$$
$$
{} - {{\xi (\varphi)}\over {m^2(\varphi)}} \left( \varphi D^\dagger _\mu
D^\mu \varphi - {{\nabla^2M(\varphi)}\over {M(\varphi)}} \varphi^2 -
4V(\varphi) - \{f \varphi \overline\psi \psi\} \right) = 0 \, , \eqno (25)
$$
where $D^\dagger _\mu$ is Hermitian conjugate of $D_\mu$, $\{f\overline
\psi\psi\}$
stands for the derivative $\partial\{f\varphi\overline\psi\psi\}/
\partial \varphi$, and
$$
\xi(\varphi) = {1\over 2} {{\partial m^2(\varphi)}\over {\partial
\varphi}} \, \eqno                       (26)
$$
or, writing explicitly the multiplet indices
$$
\xi_A (\varphi) = {1\over 2} {{\partial m^2(\varphi)}\over {\partial
\varphi^A}} = \xi_{AB}\, \varphi^B \, . \eqno (27)
$$
Note that the left-hand-side of Eq.~(25) if multiplied by  $\varphi$ with the summation
over the scalar multiplet index becomes identically zero.

For classical vacuum configurations we set $A = 0$, $\psi = \overline\psi
= 0$, $\varphi = const$, and one gets from (25) the following equation
$$
{{\partial V(\varphi)}\over {\partial \varphi}} - {{\xi(\varphi)}\over
{m^2(\varphi)}} 4 V(\varphi) = 0 \, , \eqno (28)
$$
which is simply the extremum condition of the  function  $V(\varphi)$  on the
hypersurface given by the equation $m^2 (\varphi) = const$ in the space of $\varphi$. Due
to the positivity  property  of  the  form  $m^2 (\varphi)$ this hypersurface is compact
so  Eq.~(28)  has  non-trivial  solutions. Among them there are those which minimize the
potential  $V(\varphi)$  on the hypersurface considered. The values $\varphi_0$ of this
solution  will determine in a usual way the fermionic masses through the  Yukawa coupling
terms in the Lagrangian and  the  masses  of  the  gauge vector bosons through the gauge
interactions.

Variation of the  action  (20)  over  the  fermionic  fields
yields the following equations of motion
$$
i \left( \gamma^\mu (\partial_\mu + A_\mu)\psi + {1\over 2\sqrt{-g}}
\partial_\mu (\sqrt{-g}\gamma^\mu)\psi \right) + {1\over 4}\varepsilon^a
{}_{bcd}\,e^\mu_a\,\Gamma^{bc}{}_\mu\gamma^d\gamma^5\psi \,
$$
$$
{} - \,{3\over 8m^2(\varphi)} (\overline\psi\gamma_\mu\gamma^5\psi)\gamma^\mu
\gamma^5\psi - \{f\varphi\psi\} = 0 \, . \eqno   (29)
$$

Variation of the  action  (20)  over  the  metric  with  the
equations of motion (29) taken into account yields
$$
m^2(\varphi) \left( R_{\mu\nu}(\Gamma) - {1\over 2} g_{\mu\nu}
R(\Gamma) \right) = (\nabla_\mu\nabla_\nu - g_{\mu\nu}\nabla^2)
m^2(\varphi) + D^\dagger {}_{(\mu}\varphi D_{\nu)}\varphi
$$
$$
{} - 6\partial_\mu
M(\varphi)\partial_\nu M(\varphi)
- {i\over 2} (\overline\psi\gamma
_{(\mu} D_{\nu)}\psi - \overline {(D_{(\mu}\psi)}\gamma_
{\nu)}\psi) + {1\over e^2} Tr F_{\mu\sigma}F_\nu{}^\sigma
$$
$$
{} -\, {1\over 2}\, g_{\mu\nu} \Biggl(
 |D\varphi|^2 - 6(\nabla M(\varphi))^2 + 2V(\varphi) \,
$$
$$
{} - \,  {3\over 8m^2(\varphi)} (\overline\psi\gamma^\sigma
\gamma^5\psi)(\overline\psi\gamma_\sigma\gamma^5\psi) + {1\over 2 e^2} Tr
F_{\sigma\tau}F^{\sigma\tau} \Biggr) \, .  \eqno (30)
$$

The equations obtained are invariant with respect  to  local
conformal transformations considered above. All  the  observables
of the theory are regarded to be invariant as well.  This  allows
one  to  fix  the  conformal  gauge  freedom  by  imposing   some
appropriate condition on the solutions. Two  of  such  conditions
are especially convenient as can be seen from the Lagrangian (21)
or from the equations of motion written just above. The first one
is the gauge already mentioned which is defined by the equation
$$
m^2 (\varphi) =  \mbox{const} \, . \eqno                       (31)
$$
The second is the gauge-fixing condition
$$
M^2 (\varphi) =  \mbox{const} \, . \eqno                       (32)
$$
The gauge (31) is the most convenient one  for  the  cosmological
interpretation  of  the  theory as in   this   gauge   the
gravitational coupling (or, equivalently,  the  Planck  mass)  is
explicitly constant.

\section{Cosmological scenario}

A viable cosmological  scenario  now  can  hardly  be  built without inflationary stage.
So we  start  with  the  question  of whether and under  what  conditions  does  our
model  allow  for inflation. Doing this we will have in  mind  mostly  the  chaotic
inflation scenario which  is  more  natural  and  for  which  the analysis is more simple
as compared to the new inflation  (we will study more thoroughly both these scenarios, as
they appear in our model, in our subsequent papers, for their good review see [15]). As
is usual for such an analysis, we will  assume  that  during  inflation  scalar  field
contribution dominates in the energy-momentum tensor, so the rest of the matter  fields
will  be  put  to  zero.  Considering  the standard Friedmann--Robertson--Walker
cosmology we  obtain  the equations  for  the  dynamics  of  the  universe  filled   by a
homogeneous scalar field multiplet  $\varphi (t)$.  It  is  convenient  to write these
equations in the gauge (31). In  this  gauge  we  can apply the standard  analysis  of
the  plausible  conditions  for inflation. The dynamics equations in the gauge (31)  are
written as follows
$$
H^2 + {\kappa\over a^2} = {1\over 3m^2} \left( {1\over 2} \dot
\varphi^2 - {(\varphi \dot \varphi)^2\over 12M^2} + V
\right) \, , \eqno (33)
$$
$$
\dot H - {\kappa\over a^2} = - {1\over m^2} \left( {1\over 2}
\dot \varphi^2 - {(\varphi\dot\varphi)^2\over 12M^2} \right) \, ,
\eqno (34)
$$
$$
\ddot \varphi + 3H\dot \varphi - {\ddot M + 3H\dot M\over M}
\varphi + {\partial V \over \partial \varphi}
$$
$$
{} - {\xi(\varphi)\over m^2} \left( \varphi\ddot \varphi
+ 3H\varphi\dot\varphi - {\ddot M + 3H\dot M\over M} \varphi^2 +
4V \right) = 0 \, , \eqno (35)
$$
where $H \equiv \dot a/a$ is the Hubble parameter, $\kappa = 0,\,\pm 1$
determines  the
spatial curvature of the universe,  and  the  quantity  $\xi(\varphi)$
was
defined in (26). Dots denote the derivatives with respect to  the
cosmological time $t$. It is easy to see that
$$
{1\over 2} \dot \varphi^2 - {(\varphi \dot \varphi)^2 \over
12M^2} \geq {1\over 2} \left( \dot \varphi^2 - {(\varphi \dot \varphi)^2
\over \varphi^2} \right) \geq 0 \, , \eqno (36)
$$
hence the right-hand-side of Eq.~(33)  is  not  negative  and  the right-hand-side of
Eq.~(34) is not positive.

The condition for inflation $|\dot H| \ll H^2$  implies then
$$
\dot \varphi^2 - {(\varphi\dot\varphi)^2\over 6M^2} \ll V(\varphi) \, .
\eqno (37)
$$

During inflation soon it becomes possible to neglect  the  second derivatives of the
scalar fields (the latter begin to roll slowly down the scalar field  potential  on  the
hypersurface  $m^2 (\varphi) = const$). From Eqs.~(35) and (33) we then have a very rough
estimate
$$
\dot \varphi^2 \sim {V(\varphi) \over \varphi^2} m^2 \, . \eqno (38)
$$
Taking into account the definition (7) we obtain  from  (37)  the
following both inflation and slow-rolling condition
$$
m^2(\varphi) \ll \varphi^2,$$ or $$\xi_{AB} \ll 1 \, . \eqno (39)
$$
This estimate is not difficult to  understand. The value $m^2 (\varphi)$
determines the Planck mass $M_{\rm P}$ through
$$
m^2 (\varphi)  =  {M^2_{\rm P}\over 8\pi} \, .  \eqno (40)
$$
In terms of the Planck mass the condition (39) reads
$$
M^2_{\rm P} \ll  \varphi^2 \, , \eqno           (41)
$$
and  looks  quite  familiar  to  those  who  deal  with   chaotic
inflationary cosmology (see [15]).

The estimate (39) is sufficient for inflation to take place, but not necessary. In fact,
it is the condition for the chaotic type inflation. It is clear that the condition (37)
can be fulfilled in the plateau regions of the potential $V(\varphi)$ on the hypersurface
$m^2(\varphi) = \rm{const}$ without the condition (39). In this case we would have
inflationary dynamics similar to new inflation. Concrete realization of both these
possibilities will be a subject of our future studies.

Besides (39) another condition which  is  necessary  for  the  sufficient
amount of inflation to take place is the following
$$
\lambda_{ABCD} \ll 1 \, . \eqno    (42)
$$
It stems from (41) and from the requirement $V(\varphi) \lesssim (m^2 ( \varphi))^2$
which allows one to consider space-time as classical.

Neglecting terms with the  second  time  derivative  of  the scalar fields $\varphi$ in
Eq.~(35) and using the condition  (39)  we  are able to write down the following
approximate equation for $\varphi$
$$
3H \left(\dot \varphi - {\varphi\dot\varphi \over \varphi^2} \varphi \right)
+ {\partial V(\varphi)\over \partial \varphi} - {\xi(\varphi)\over
m^2(\varphi)} 4V(\varphi) = 0 \, . \eqno (43)
$$

Note that if multiplied by $\varphi$ with  the  summation  over  the
scalar multiplet index this equation gives identically  zero,  as
is also the case with the precise equation of motion for $\varphi$.

The  slow-rolling  regime  of  the  scalar  field   dynamics
terminates as the scalar field values approach close to a (may be
local) minimum of the  potential $V(\varphi)$  on  the  hypersurface  of
constant $m^2 (\varphi)$. After that the scalar  fields  start  oscillating
around their stationary point. Due to various  couplings  between
fields these oscillations give  birth  to  particles  of  various
kinds. This process heats the universe in the standard manner.

The value $V_0$  of the scalar field potential in  its  local minimum (on the
hypersurface $m^2 (\varphi) = \mbox{const}$) generates the present-day $\Lambda$-term
whose value can be estimated as $\Lambda \sim (\lambda/ \xi^2) M^4_{\rm P}$ (here
$\lambda$ has the order of magnitude of the $\lambda$'s in (8), and $\xi$ has the order
of magnitude of the $\xi$'s in (7)). In general this $\Lambda$-term is nonzero. For this
not  to  be the case  some fine tuning  of  the  potential  parameters $\lambda_{ABCD}$
seems  to  be necessary. If we assume the scalar field potential $V(\varphi)$ to  be
non-negative then to have $V_0  = 0$ in the minimum it is necessary and sufficient  that
such $\varphi_0 \not= 0$ exists for which $V(\varphi_0) = 0$. An interesting suggestion
is that this property somehow can be  provided  by  other  symmetries  of  the theory
(besides the local conformal one).

\section{Conformally invariant quantum cosmology}

In this section our  aim  will  be  to  see  how  the standard formalism of quantum
cosmology  (for  reviews  see  [16--19])  is applied to the conformally invariant theory
which  we  study  in this paper. For simplicity we will restrict ourselves only to the
scalar-gravitational sector of the full  theory  (see  Eqs.~(20) and (21) for the
action), which is described by the Lagrangian
$$
L  = {1\over 2}m^2(\varphi)R(\Gamma) - {1\over 2} \left(\nabla\varphi
\right)^2 + 3\left(\nabla M(\varphi)\right)^2 - V(\varphi) \, , \eqno
(44)
$$
and by the action
$$
S  = \int \limits_M L\sqrt{-g} d^4x + \int\limits_{\partial M}
m^2 (\varphi) {\cal K}(\Gamma)\sqrt h d^3x \, , \eqno (45)
$$
in which the boundary term  was  explained  in  Section  2  after Eq.~(21).

The metric $g$  in  the  ADM  form  (see  [16]  for  the  best
description) is written as
$$
g = -\left( N^2 - N_i N^i\right) dt\otimes dt + N_i (dt\otimes dx^i
+ dx^i\otimes dt) + h_{ij} dx^i \otimes dx^j \, . \eqno (46)
$$
Latin indices from the middle of the alphabet run from  1  to  3.
The quantities $N$ and $N_i$  are respectively  lapse  function
and shift vector. Substituting the metric in this form  into  the
action (45) we obtain for a spatially closed universe
$$
S  = \int\limits_M {\cal L}\,d^3x\,dt \, , \eqno            (47)
$$
where the Lagrangian density ${\cal L}$ is given by
$$
{\cal L} = {1\over 2} N\sqrt h \,  m^2(\varphi) \left( {\cal R} +
{\cal K}_{ij}{\cal K}^{ij} - {\cal K}^2   \right)
$$
$$
{} + N\sqrt h \left( {2 {\cal K}
\over N} \xi_A(\varphi)\dot\varphi^A  - \left(m^2 (\varphi)\right)^{|i}
{}_{|i} - {2{\cal K}N^i \over N} \xi_A(\varphi)\partial_i\varphi^A - V(\varphi)
\right)
$$
$$
{} + N\sqrt h \, T_{AB}(\varphi) \left( {1\over 2N^2} \dot \varphi^A \dot
\varphi^B
- {N^i\over N^2} \dot \varphi^A\partial_i\varphi^B
        - {1\over 2} \left(
h^{ij} - {N^iN^j \over N^2} \right)\partial_i\varphi^A\partial_j
\varphi^B   \right) \, , \eqno (48)
$$
and
$$
{\cal K}_{ij} = {1\over 2N} \left( 2N_{(i|j)} - \dot h_{ij}
     \right) \eqno (49)
$$
is the extrinsic curvature of the hypersurface $t = \mbox{const}$  in  the ADM  form.
Vertical lines  denote  covariant  derivatives  with respect to the metric $h_{ij}$, with
the aid of which also small Latin indices are raised and lowered, ${\cal R}$ is the
three-curvature  scalar built of the metric $h_{ij}$, and
$$
T_{AB}(\varphi) = \delta_{AB} - {6\over M^2(\varphi)} \left(
\xi_A(\varphi) + {1\over 6} \varphi^A \right) \left(\xi_B(\varphi)
+ {1\over 6} \varphi^B \right) \eqno (50)
$$
is a non-degenerate symmetric matrix of signature $(-,+,+,...,+)$.

The theory described by the Lagrangian (48)  is  degenerate:
due  to  coordinate   reparametrization   invariance   the   time
derivatives (velocities) of the values $N$ and $N_i$  do not enter  the
Lagrangian at all; due to local conformal invariance  the  matrix
of the second derivatives of ${\cal L}$ with respect to the velocities
$\dot h_{ij}$
and $\dot \varphi^A$  has one zero eigenvalue. So some linear combination
of the
velocities  $\dot h_{ij}$ and $\dot \varphi^A$   will  not  be
expressed  through   the
corresponding  generalized  momenta.   As   a   consequence   new
constraint will appear.

Let us denote the generalized momenta for the variables $h_{ij}$
and $\varphi^A$  by $p^{ij}$ and $p_A$  respectively. Then given
the Lagrangian (48)
we can  express  the  velocities $\dot \varphi^A$ through  the
corresponding momenta
$$
\dot \varphi^A = N^i\partial_i \varphi^A + {N\over \sqrt h} \,
T^{AB}(\varphi) \left( p_B - 2\sqrt h \,{\cal K} \xi_B(\varphi)
\right) \, , \eqno (51)
$$
where $T^{AB}(\varphi)$ is the matrix inverse  of  $T_{AB}(\varphi)$.
Trying  then  to
express the velocities $\dot h_{ij}$, we find that this  is
possible  only for the traceless part of the tensor $\dot h_{ij}$. We have
$$
\widetilde {\cal K}^{ij} = - {2\over m^2(\varphi)\sqrt h} \widetilde p^{ij}
\, , \eqno (52)
$$
where $\widetilde {\cal K}^{ij}$ and $\widetilde p^{ij}$ are the traceless
parts
of ${\cal K}^{ij}$ and $p^{ij}$, respectively,
the values ${\cal K}_{ij}$ being given by  (49).
The  trace $ p  =  p^{ij} h_{ij}$    is
involved into the constraint
$$
{\cal F} \equiv {1\over 3} \left( p^{ij} h_{ij} - {1\over 2}
p_A\varphi^A \right) =  0 \, . \eqno                 (53)
$$

The action written in the Hamiltonian form is
$$
S = \int\limits_M \left( p_A\dot\varphi^A + p^{ij} \dot h_{ij}
- v{\cal F} - N{\cal H} - N^i{\cal H}_i \right) d^3xdt \, .
\eqno  (54)
$$
Variables $v$, $N$, $N^i$   are  Lagrange  multipliers.  Variations
  with
respect to them give constraint equations. The expression  for
 ${\cal F}$
is written in (53). Expressions for ${\cal H}$ and ${\cal H}_i$  are
$$
{\cal H} = {2\over \sqrt hm^2(\varphi)} \widetilde p^{ij} \widetilde p_{ij}
 + {1\over 2\sqrt h}\, T^{AB}(\varphi)\,p_Ap_B - {1\over 2} m^2(\varphi)
\sqrt h \,{\cal R}
$$
$$
{} + \sqrt h \left(m^2(\varphi)\right)^{|i}{}_{|i} + {1\over 2}
\sqrt h \, h^{ij}T_{AB}(\varphi) \partial_i\varphi^A\partial_j
\varphi^B + \sqrt h \, V(\varphi) \, , \eqno (55)
$$
$$
{\cal H}_i = - {1\over 3} \left( \varphi^Ap_A\right)_{|i} -
2\widetilde p^j_{i|j} + \partial_i\varphi^Ap_A \, . \eqno (56)
$$
Note that the trace of $p^{ij}$   does not enter the constraints
${\cal H}$  and ${\cal H}_i$. To simplify the equations of the theory
 it is  convenient  to
make  a  canonical  transformation  to  new  canonical  variables
related to the old ones through the following equations
$$
\widetilde h_{ij} = h^{-1/3}h_{ij}, \,\, \widetilde \pi^{ij} =
h^{1/3} \widetilde p^{ij} \, , \eqno (57)
$$
$$
h = det(h_{ij}), \,\, \pi_h = {1\over 3h} \left( p^{ij}h_{ij}
- {1\over 2} p_A\varphi^A \right) = {1\over h} {\cal F} \, ,
\eqno (58)
$$
$$
\chi^A = h^{1/6}\varphi^A, \,\, \pi_A = h^{-1/6} p_A \, .
\eqno (59)
$$
The meaning of the first line (57) in the above relations is  the
following: $\widetilde h_{ij}$ is  a  function  of  some  five  parameters
(not
specified here) which determine this matrix with  unitary  trace,
$\widetilde \pi^{ij}$ is a function of the corresponding five generalized
momenta.
Denoting also $h^{-1/6}N = \widetilde N$, $h^{1/6}{\cal H} = \widetilde
{\cal H}$, we will have $N{\cal H} = \widetilde N\widetilde {\cal H}$, and
$$
S = \int \limits_M \left( \pi_A\dot\chi^A + \widetilde\pi^{ij}\dot{\widetilde
h}_{ij} - (v - {\dot h \over h} ){\cal F} - \widetilde N\widetilde {\cal H}
- N^i {\cal H}_i \right) d^3x\,dt\, , \eqno (60)
$$
where in terms of new variables
$$
{\cal F} = h\pi_h\, , \eqno (61)
$$
$$
\widetilde {\cal H} = {2\over m^2(\chi)} \widetilde \pi^{ij}\widetilde
\pi_{ij} + {1\over 2} T^{AB}(\chi) \pi_A\pi_B - {1\over 2} m^2
(\chi) \widetilde {\cal R}
$$
$$
{} + \left( m^2(\chi) \right) ^{\cdot i}{}_{\cdot i} + {1\over 2} \widetilde
h^{ij}
T_{AB} (\chi) \partial_i\chi^A\partial_j\chi^B + V(\chi)\, , \eqno (62)
$$
$$
{\cal H}_i = - {1\over 3} \left( \chi^A\pi_A \right)_{\cdot i} -
2 \widetilde \pi^j_{i \cdot j} + \partial_i\chi^A\pi_A \, . \eqno (63)
$$
Dots in front of small Latin indices denote covariant derivatives
with respect to the metric $\widetilde h_{ij}$, $\widetilde {\cal R}$
is the curvature  scalar  built
of this metric. All small Latin indices in (62), (63)  and  below
are also lowered and raised by the metric $\widetilde h_{ij}$
and  its  inverse $\widetilde h^{ij}$.

{}From the expression (60) for the  action  we  see  that  the
variables $h$, $\pi_h$  are non-dynamical. Their role becomes manifest if
we write down a  formal  quantum  path  integral  of  the  theory
described by the action (60). For the  wave  function $\Psi$  of  the
universe we will have
$$
\Psi = \int e^{iS}\,[d\widetilde N]\,[dN_i]\,[dv]\,[d\pi_h]\,[dh]\,
[d\widetilde \pi^{ij}]\,[d\widetilde h_{ij}]\,[d\pi_A]\,[d\chi^A]\, ,\eqno (64)
$$
In this path integral it is convenient to shift  the  integration
variable $v \rightarrow v + {\dot h \over h}$
(see the expression (60)  for  the  action).
Then the integrals over the (shifted) Lagrange multiplier $v$  and
over the canonical variables $\pi_h$ and $h$  result  in  the  overall
factor
$$
\int e^{-i\int v{\cal F} d^3x\,dt}\, [dv] \,[d\pi_h]\,[dh] = \int \delta
[{\cal F}]\,[d\pi_h]\,[dh] = \int [d\sigma]\, , \eqno (65)
$$
where by $\sigma$ we denoted the integration variable $\ln h$.
In  deriving
(65) the expression (61) for the constraint ${\cal F}$ has been taken into
account. The last path integral over $\sigma$ is just the integral  over
the gauge conformal group. We see that  this  integral  has  been
factorized automatically.

The quantum  operator  constraint $\widehat {\cal F}$  imposed on  the  wave function of
the universe with the operator  arrangement as written in Eq.~(61) implies that the wave
function  does  not depend on the variable $h$. Note that the remaining variables
$\widetilde h_{ij}$, $\chi^A$  are conformally  invariant.  Thus  the  additional
constraint implies that  the  wave  function  depends  only  on  conformally invariant
combinations of the initial variables, that is,  it  is conformally  invariant.  The
constraint $\widetilde {\cal H}_i$ as   usual   means invariance of the wave function
with respect to three-metric  and matter field variations induced  by  coordinate
diffeomorphisms. All the dynamical content of the theory considered  is  expressed by the
analogue of the Wheeler--DeWitt equation
$$
\widehat {\widetilde {\cal H}} \Psi = 0 \, . \eqno (66)
$$

It  might  seem  that  by  elimination  of  the  conformally non-invariant variable $h$
from the  equations  of  the  theory  we succeeded in building a consistent conformally
invariant  quantum gravity. This would be the case if  the  theory  defined  by  the
constraint  equation  (66)  were  well-defined.  But  the  latter condition is not true
in  our  theory  as  well  as  the  similar statement is not true in the  Einstein
theory  of  gravity.  The reason is the infinity of the number of the  degrees  of
freedom together with high non-linearity of the theory. By  all  this  we are led to the
necessity of regularizations. Nevertheless  it  is interesting to note that the theory
developed here is no  "worse" in this respect  than  the  usual  Einstein  theory  of
gravity. Ascribing one or the  other  meaning  to  Eq.~(66)  (for  example, restricting
it  to  minisuperspace)  we  obtain  a  well-defined theory. Only in our case we are to
respect the possible  remnants of both the coordinate  reparametrization  invariance  and
local conformal invariance. In the following section we will consider quantum theory of
minisuperspace based on  our  model,  which  is invariant  with  respect  to  spatially
homogeneous   conformal transformations.

The basic features of the constraint $\widetilde {\cal H}$ resemble those of the
well-known Wheeler--DeWitt constraint in the  Einstein  theory  of gravity. The quadratic
form in momenta in (62) is  non-degenerate and has eigenvalue signs $(-,+,...,+)$ and the
potential part of $\widetilde {\cal H}$ is not bounded from below. In our theory  however
negative  sign comes from the matter (not metric) kinetic term in the expression (62) for
$\widetilde {\cal H}$. Pseudo-Euclidean superspace metric signature  enables us to have a
timelike variable as in standard quantum  cosmology. Such variable can be taken to be
proportional to $m(\chi)$.

To introduce time explicitly it is convenient once again  to
proceed to new variables
$$
\mu = \sqrt 6 \, m(\chi)\, , \,\, \vartheta^r = \vartheta^r(\chi)\, ,
\, r = 1,\ldots ,\,k-1 \, , \eqno (67)
$$
with corresponding conjugate momenta $\pi_\mu, \, \pi^\vartheta_r, \, r = 1,\ldots,k-1$.
If coordinates $\vartheta^r$ are chosen so that $\vartheta^r = const$ are  rays  in  the
$\chi$-space which begin at the origin $\chi = 0$, then the quadratic forms in $\pi_\mu$
and $\pi^\vartheta_r$ decouple in $\widetilde {\cal H}$ and the matter part of the
constraint $\widetilde {\cal H}$ takes the following form:
$$
\widetilde {\cal H}_{\rm matter} = - {1\over 2} \pi^2_\mu - \mu_{\cdot i}
\mu^{\cdot i} + {1\over 6} (\mu^2)^{\cdot i}{}_{\cdot i} + {1\over 2\mu^2}
E^{rs}(\vartheta)\pi^\vartheta_r\pi^\vartheta_s
$$
$$
{} + {1\over 6} \left( {\mu\over m(n)} \right)^2 \left( n^A_{\cdot i}
n_B^{\cdot i} + {1\over M^2(n)} m(n)_{\cdot i} m(n)^{\cdot i} \right) +
{1\over 2} \mu^4W(\vartheta) \, , \eqno (68)
$$
where $E^{rs}(\vartheta)$, $r,\, s = 1,\ldots,k-1$, is the symmetric  matrix  of  some
positive-definite form, $n^A = n^A(\vartheta)$, $A = 1,\ldots,k$, is a unit  vector in
the $\chi$-space as a function of $\vartheta$, and
$$
W(\vartheta) = {1\over 18m^4(n(\vartheta))} V(n(\vartheta))
\eqno (69)
$$
is the (rescaled) scalar field potential on the  hypersurface $m^2(\chi) = \mbox{const}.$

\section{Conformally invariant quantum minisuperspace cosmology}

In the  previous  section  we  were  able  to  see  how  the
well-developed ADM formalism of quantum cosmology is  applied  to
the conformally invariant theory described above. For  simplicity
we restricted ourselves only to the scalar - gravitational sector
of the full theory, which is described  by  the  Lagrangian  (44)
with the action (45). In the previous section a  theory  of  full
superspace was considered. For many purposes however,  especially
when one deals with quantum cosmology, it is desirable to perform
a more simple analysis of a minisuperspace model.  This  will  be
our main task in what follows.

For  definiteness  we  start from  homogeneous  but  not necessarily isotropic closed
cosmology known as  Bianchi-IX.  The metric for this model can be written as (see, e.g.,
[19])
$$
ds^2 = {1\over 2\pi^2} \left( - N^2(t)dt\otimes dt + e^{2\alpha (t)}
\left( e^{2\beta (t)} \right) _{ij} \sigma^i\otimes\sigma^j \right) \, ,
\eqno (70)
$$
where $N(t)$ is the lapse function, $\{\sigma^i\}$
is a homogeneous basis  of
one-forms on a unit spatial  three-sphere, $e^{\alpha (t)}$
is  the  scale
factor of the space parametrized by the function $\alpha (t)$
while  the
symmetric traceless matrix $\beta_{ij} (t)$
parametrizes  the  anisotropy.
The shift vector is absent due  to  spatial  homogeneity  of  the
metric. The overall factor out front is a convenient scaling. The
traceless matrix  $\beta_{ij}$
may  be  chosen  to  be  diagonal.  It  is
convenient to write it in the form
$$
\beta_{ij} = {1\over \sqrt 6} diag \left( \beta_+ + \sqrt 3 \beta_-,
\, \, \beta_+ - \sqrt 3 \beta_-, \, \, - 2\beta_+ \right) \, . \eqno (71)
$$

For the metric (70)  and  for  a  homogeneous  scalar  field
multiplet $\varphi (t)$ the action (45),
after taking  integral  over  the space, is written as
$$
S = \int { \cal L} \, dt \, , \eqno (72)
$$
with the Lagrangian
$$
{\cal L} =  {e^{3\alpha}\over 2N} m^2(\varphi) \left(
\dot \beta^2_+ + \dot \beta^2_-  - 6\dot\alpha^2 + T_{AB}
(\varphi) \dot\varphi^A\dot \varphi^B - 12\dot\alpha\xi_A(\varphi)
\dot\varphi^A \right)
$$
$$
{} + Ne^\alpha m^2(\varphi)R_{\rm IX} (\beta) - Ne^{3\alpha}V(\varphi)
\, , \eqno (73)
$$
in which the potential $V(\varphi)$ has been rescaled by $2\pi^2$ for
convenience, and
$$
R_{\rm IX} (\beta) = Tr \left( 2e^{-2\beta} - e^{4\beta} \right)
\eqno (74)
$$
is half of the spatial three-curvature scalar of the unitary
$(\alpha = 0)$ Bianchi-IX  space.
The  non-degenerate  matrix  $T_{AB} (\varphi)$  was
defined in (50) and the value $\xi_A (\varphi)$ - in (27).

The theory described by the Lagrangian (73)  is  degenerate:
due to time  reparametrization  invariance  the  time  derivative
(velocity) of the lapse function $N$ does not enter the  Lagrangian
at all; due to the remnant of the conformal invariance the matrix
of the second derivatives of ${\cal L}$
with respect to the  velocities  $\dot \alpha$
and $\dot \varphi$  has one zero eigenvalue.
So,  as  we  already  know,  some
linear combination of  the  velocities  $\dot \alpha$  and  $\dot \varphi$
will  not  be
expressed through the corresponding  generalized  momenta.  As  a
consequence new constraint will appear.

Let us denote the generalized momenta for the variables $\varphi^A, \,
\alpha,\, \beta_+$,  and $\beta_-$  by $p_A, \, p, \, p_+$, and $p_-$
respectively. Then  given  the
Lagrangian (73) we can express the velocities $\dot\varphi^A$
and $\dot \beta_\pm$   through
the corresponding momenta
$$
\dot \varphi^A = T^{AB} (\varphi) \left( Ne^{-3\alpha}p_B +
6\dot\alpha\xi_B(\varphi) \right) \, , \eqno (75)
$$
$$
\dot \beta_\pm = {N\over m^2(\varphi)} e^{-3\alpha}p_\pm
\, , \eqno (76)
$$
where $T^{AB} (\varphi)$ is the matrix inverse  of  $T_{AB} (\varphi)$.
Trying  then  to
express the velocity $\dot\alpha$ through the momenta we find that  this  is
not possible. The value $p$ turns to be involved in the constraint
$$
{\cal F} (\varphi^A,p_A,\alpha,p) \equiv p - \varphi^Ap_A = 0
\, . \eqno (77)
$$

The action written in the Hamiltonian form is
$$
S = \int \ \left( p_A\dot\varphi^A + p_+\dot\beta_+ +
p_-\dot\beta_- + p\dot\alpha - v {\cal F} - N {\cal H} \right) dt
\, . \eqno (78)
$$
The variables $v$ and $N$, are the Lagrange  multipliers.  Variations
with respect to them give constraint  equations.  The  expression
for ${\cal F}$ is written above in (77). The expression for ${\cal H}$ is
$$
{\cal H} = e^{-3\alpha} \left( {1\over 2m^2(\varphi)} \left( p^2_+ +
p^2_- \right) + {1\over 2} T^{AB} (\varphi)p_Ap_B \right) - e^\alpha
m^2(\varphi)R_{\rm IX} (\beta) + e^{3\alpha}V(\varphi) \, . \eqno (79)
$$
Note that the canonical variable $p$ does not enter the  constraint
${\cal H}$. To simplify the equations of the theory it  is  convenient
to make a canonical  transformation  from
$\varphi^A,\, p_A$,  and  $p$  to  new
canonical  variables $\chi^A,\, \pi_A$,  and $\pi_\alpha$ ($\alpha,\,
\beta_\pm$,  and  $p_\pm$   being
untouched)  related  to  the  old  ones  through  the   following
equations (compare with (57)-(59))
$$
\chi^A = e^\alpha\varphi^A, \,\,\, \pi_A = e^{-\alpha}p_A, \, \,\,
\pi_\alpha = {\cal F} (\varphi^A,p_A,\alpha,p) = p - \varphi^Ap_A
\, . \eqno (80)
$$
Denoting also $e^{-\alpha}N = \widetilde N, \, e^\alpha {\cal H} =
\widetilde {\cal H}$, we will have  $N {\cal H}  = \widetilde N \widetilde
{\cal H}$, and
$$
S = \int \ \left( \pi_A\dot\chi^A + p_+\dot\beta_+ +
p_-\dot \beta_- - (v - \dot \alpha) \pi_\alpha - \widetilde N
\widetilde {\cal H} \right) dt \, , \eqno (81)
$$
where in terms of new variables
$$
\widetilde {\cal H} = {1\over 2m^2(\chi)} \left( p^2_+ + p^2_- \right)
+ {1\over 2} T^{AB} (\chi) \pi_A\pi_B - m^2(\chi)R_{\rm IX} (\beta)
+ V(\chi) \, . \eqno (82)
$$

{}From the expression for the action  (81)  we  see  that  the
variables $\alpha$ and $\pi_\alpha$  are non-dynamical.
Their role becomes manifest
if we write down the formal quantum path integral of  the  theory
described by the action (81). For the wave function
$\Psi (\alpha,\beta,\chi^A)$  we have
$$
\Psi = \int e^{iS} \, [d\widetilde N] \, [dv] \, [d\pi_\alpha] \,
[d\alpha] \, [dp_\pm] \, [d\beta_\pm] \, [d\pi_A] \, [d\chi^A]
\, , \eqno (83)
$$
After shifting the integration variable $v \rightarrow v + \dot\alpha$
(see  the
expression (81) for the  action)  the    integrals    over    the
(shifted) Lagrange multiplier $v$ and over the canonical  variables
$\pi_\alpha$ and $\alpha$ lead to an overall factor
$$
\int e^{-i\int v\pi_\alpha dt} \, [dv] \, [d\pi_\alpha] \,
[d\alpha] = \int \delta [\pi_\alpha] \, [d\pi_\alpha] \,
[d\alpha] = \int \, [d\alpha] \, . \eqno (84)
$$
The last path integral over $\alpha (t)$
is just the (infinite)  integral
over the remnant of the local conformal group. It is
the analogue of the factor (65) of the theory of full superspace.
Again we see that this integral factorizes automatically.

Quantum operator constraint $\widehat \pi_\alpha$ imposed on the wave  function of the
universe implies that the wave function $\Psi (\alpha,\beta,\chi^A)$ does not depend on
the variable $\alpha$. Note that the remaining  variables $\beta_\pm, \, \chi$  are
conformally  invariant. Thus  the  additional  constraint implies that  the  wave
function  depends  only  on  conformally invariant combinations of initial variables,
that is, that it  is conformally invariant. All the dynamical content of the theory is
then expressed by the analogue of the Wheeler-DeWitt equation for minisuperspace  (see
[15--19]  for  a  review  of  the   standard minisuperspace quantum cosmology)
$$
\widehat {\widetilde {\cal H}} = 0 \, . \eqno (85)
$$
In imposing constraints on  the  wave  function  some  particular
operator ordering is to be chosen.  Different  choices  as  usual
correspond  to  different  quantum  versions  of  the   principal
classical theory. We will not discuss this topic here.

The basic features of the constraint $\widetilde {\cal H}$  were described  in general in
the previous  section.  They  resemble  those  of  the well-known Wheeler-DeWitt
constraint in the  Einstein  theory  of gravity (see [15--19]). The quadratic form in
momenta in  (82)  is non-degenerate and has  eigenvalue  signs $(-,+,\ldots,+)$.
Negative sign comes from the matter kinetic term in  the  expression  (82) for
$\widetilde {\cal H}$ and enables us to have a timelike variable as  in  standard
minisuperspace quantum cosmology.

To introduce time explicitly it is convenient to proceed  to new variables (67) with
corresponding conjugate momenta $\pi_\mu, \, \pi^r_\vartheta, \, r = 1,\ldots,k-1$. If
the  coordinates  $\vartheta^r$   are  chosen  as  in  the previous section  (see  the
text  following  Eq.~(67))  then  the quadratic forms in $\pi_\mu$  and $\pi^\vartheta_r$
decouple in  the  expression  for  $\widetilde {\cal H}$ which then takes the following
shape
$$
\widetilde {\cal H} = {1\over 2\mu^2} \left( p^2_+ + p^2_- \right)
- {1\over 2} \pi^2_\mu + {1\over 2\mu^2} E^{rs} (\vartheta)
\pi^\vartheta_r\pi^\vartheta_s - {1\over 6} \mu^2R_{\rm IX} (\beta)
+ {1\over 2} \mu^4 W (\vartheta) \, , \eqno (86)
$$
where $E^{rs} (\vartheta), \, r,s = 1,\ldots,k-1$,
is the matrix  of  the same  positive-definite form,
as in (68), and $W(\vartheta)$ is given by (69).

{}From the description  given  at  the  end  of  the  previous
section it is clear that $\{\vartheta\}$ can be regarded as
just  (arbitrary)
coordinates on the unit $(k-1)$-sphere  in  the  $\chi$-space.
This  in
particular has the following consequence. If $\mu$  is  a  good  time
variable (the wave function of the universe has WKB  form  in  $\mu$)
then the matter field probability is distributed over a  manifold
with the topology of $(k-1)$-sphere. This manifold is compact hence
the potential $W(\vartheta)$ given  by  (69)  is  bounded  on  it
and  the
probability distribution may be well defined on it everywhere. To
illustrate this idea let us  consider  the  case  when  the  wave
function has  a  WKB  form  in  the  time  variable  $\mu$  and,  for
simplicity, let us ignore the variables $\beta_\pm$
i.e. put $\beta_\pm  = 0$  (this
means that we turn to the de Sitter minisuperspace model). Let  us
also assume that the action of the third term in (86) on the wave
function is negligible. The effective constraint operator is then
$$
{\cal H}_{\rm eff} = - {1\over 2} \left( \pi^2_\mu + \mu^2
- \mu^4 W(\vartheta) \right) \, , \eqno (87)
$$
and the wave function is written in the WKB form as
$$
\Psi (\mu , \vartheta) = A(\vartheta) \exp \left( I(\mu, \vartheta)
\right) \, , \eqno (88)
$$
where $A(\vartheta)$ is the normalization factor. Hence
$\vartheta$  (in  fact  $W(\vartheta)$)
play the role of parameters in the equation
$$
\widehat {\cal H}_{\rm eff} \Psi = 0 \, . \eqno (89)
$$
In the leading order of the WKB approximation  the  solution  for
$I(\mu,\vartheta)$ is
$$
I(\mu,\vartheta) = \cases { \pm {1\over 3W(\vartheta)} \left(
1 - \mu^2W(\vartheta) \right)^{3/2}, & $\mu^2 W(\vartheta)
\leq 1 $; \cr
\pm {i\over 3 W(\vartheta)} \left(\mu^2W(\vartheta) - 1 \right)
^{3/2}, & $\mu^2 W(\vartheta) \geq 1 \,$. \cr} \eqno (90)
$$
In  order  to  be  able  to  neglect  the  action  of  the   term
$(1/ 2\mu^2) E^{rs} (\vartheta) \pi^\vartheta_r\pi^\vartheta_s$
on the wave function the normalization factor $A(\vartheta)$
is to be taken
$$
A(\vartheta) = \exp \left( \pm {1\over 3 W(\vartheta)} \right)
\, . \eqno (91)
$$
The WKB solution (88) is then valid for not very large values  of
$\mu$. The signs in (90) and (91)  are  determined  by  the  boundary
conditions. Typically the wave function is a  linear  combination
of the exponents of (90) with different signs.

Let us consider, as  an  example,  two  classical  choices  of the boundary conditions
and the corresponding wave functions. The tunneling boundary condition of  Vilenkin
[20--22]  demands  that there be only an outgoing wave at the singular boundary $\mu
\rightarrow \infty$  of the  minisuperspace.  The  corresponding  wave  function  in  the
leading order of the WKB approximation is given by
$$
\Psi_T(\mu,\vartheta) = \cases { \exp \left( - {1\over 3 W(\vartheta)}
\left[ 1 - \left( 1 - \mu^2 W(\vartheta) \right)^{3/2} \right]
\right), & $\mu^2 W(\vartheta) \leq 1 $; \cr
\exp \left( - {1\over 3 W(\vartheta)} \left[ 1+ i \left( \mu^2
W(\vartheta) - 1 \right)^{3/2} \right] \right), & $\mu^2W(\vartheta)
\geq 1.$ \cr} \eqno (92)
$$

The  boundary  condition  of  Hartle  and  Hawking   [23--26]   is formulated in terms of
Euclidean path integral representation  of the wave function similar to (83). The
proposal is that the  wave function  is  given   by   the   path   integral   over
compact configurations without the second boundary. Application  of  this proposal to the
model considered here can be performed just along the same lines as it is done for the
minisuperspace  model  based on the Einstein theory of gravity [27]. We obtain  the
following result:
$$
\Psi_H(\mu,\vartheta) = \cases { \exp \left( {1\over 3 W(\vartheta)}
\left[ 1 - \left( 1 - \mu^2 W(\vartheta) \right)^{3/2} \right]
\right), & $\mu^2 W(\vartheta) \leq 1 $; \cr
\exp \left( {1\over 3 W(\vartheta)}\right) \cos \left[ {1\over 3W(\vartheta)}
\left( \mu^2W(\vartheta) - 1 \right)^{3/2} - {\pi\over 4} \right], &
$\mu^2W(\vartheta) \geq 1 $. \cr} \eqno (93)
$$

According to  the  widespread  interpretation  of  the  wave
function [28] the probability density is to  be  defined  through
the probability flux vector on any of the  hypersurfaces  in  the
minisuperspace which is crossed by  the  probability  flux  lines
from the same side everywhere. In our case such a hypersurface is most
conveniently chosen as $\mu = const$. The probability density is then
defined in the space of variables $\{\vartheta\}$ and is given by
$$
dP(\vartheta) = J^{(\mu)} (\vartheta) d\Sigma_{(\mu)} (\vartheta)
\, , \eqno (94)
$$
where
$$
J^{(\mu)} = Im \left( \Psi {\partial\over \partial\mu} \Psi^* \right)
\eqno (95)
$$
is the probability flux vector component in the direction  of $\mu$,
and $d\Sigma_{(\mu)} (\vartheta)$ is the surface volume element of
the  surface  $\mu = const$.
This volume element is easily calculable to be
$$
d\Sigma_{(\mu)} = {\mu^{k-1}\over m^k ( n (\vartheta))}\, \sqrt {
{m^2(n(\vartheta)) + 6 \xi^2(n(\vartheta)) \over m^2(n(\vartheta))
+ {1\over 6} }}\, dS^{k-1} (\vartheta) \, , \eqno (96)
$$
where $\xi^2 = \sum_A \xi_A\xi_A$, $k$  is  the  dimension  of  the
$\chi$-space,  and
$dS^{k-1}(\vartheta)$ is the surface volume element of the unit
sphere $S^{k-1}$    in
the $\chi$-space. Calculating the fluxes
$J_T{}^{(\mu)}$ and $J_H{}^{(\mu)}$    correspondingly
for the tunneling wave function of  Vilenkin  (92)  and  for  the
expanding-universe component of the wave function of  Hartle  and
Hawking (93)  we  obtain  the  expressions  for  the  probability
densities in the corresponding cases $(\mu^2 W > 1)$
$$
dP_T(\vartheta) = C_T \sqrt { \mu^2W(\vartheta) - 1 }\,
\exp \left( - {2\over 3 W(\vartheta)} \right) \, \left(
{\mu\over m(n(\vartheta))} \right)^k
$$
$$
{} \times \sqrt {
{m^2(n(\vartheta)) + 6 \xi^2(n(\vartheta)) \over m^2(n(\vartheta))
+ {1\over 6} }}\, dS^{k-1} (\vartheta) \, , \eqno (97)
$$
$$
dP_H(\vartheta) = C_H \sqrt { \mu^2W(\vartheta) - 1 }\,
\exp \left( {2\over 3 W(\vartheta)} \right) \, \left(
{\mu\over m(n(\vartheta))} \right)^k
$$
$$
{} \times \sqrt {
{m^2(n(\vartheta)) + 6 \xi^2(n(\vartheta)) \over m^2(n(\vartheta))
+ {1\over 6} } }\, dS^{k-1} (\vartheta) \, , \eqno (98)
$$
where $C_T$  and $C_H$  are corresponding  normalization  constants.
The
expressions obtained differ only in the signs in  the  powers  of
the leading exponent factors. Thus  for  the  tunneling  boundary
conditions the probability to have large potential $W(\vartheta)$
at  the
onset of classical  universe  evolution  is  exponentially  large
whereas for the boundary conditions of Hartle  and  Hawking  such
probability  is  exponentially  suppressed.   This   means   high
probability of inflation in the first case and low in the  second
one.

To conclude this section we wish to stress once  again  that
the probabilities (97) and (98)  are  distributed  on  a  compact
manifold with the topology of $(k-1)$-sphere in the  space  of  the
fields $\{\chi\}$. On this manifold the field potential
$W(\vartheta)$ is bounded,
and if it is also smooth enough then the expressions  like  (97),
(98) may be applicable everywhere on this  manifold.  This
situation is much different from what we have in usual cases when
probability distributions are defined typically on spaces like
$R^n$ with the scalar field potential unbounded.

\section{Discussion}

In this paper we presented a generalization  of  a  GUT-like model which incorporated
gravity and which  was  invariant  under the group of local conformal transformations.
The model was based on Riemann--Cartan geometry and the vector trace of torsion played
the role of gauge vector potential for the  conformal  group.  We have considered
inflationary universe dynamics based on our model and found that  inflationary  stage  is
allowable  provided  the couplings satisfy some natural  constraints.  We  also developed
standard  quantum  gravitational  formalism  for  the  scalar-gravitational sector of the
model. We have seen that  conformally non-invariant dynamical variables can  be
eliminated  from  the equations of the theory so that the wave function of the universe
turns to be independent of them. Although our treatment was rather  formal, nevertheless,
it  may  indicate  that  it  is possible to construct  and  operate  with conformally
invariant quantum theory of gravity.

We also considered a simple  minisuperspace  formulation  of
the  theory  under  discussion.  It  has  been  illustrated  that
invariance with respect to local conformal transformations (to be
precise, their spatially homogeneous remnant) can be consistently
implemented  into  a  quantum  theory  of  minisuperspace  by  an
appropriate  operator  arrangement  choice.  Then  this  symmetry
becomes manifest as conformal invariance of  the  wave  function,
i.e. its dependence only on conformally invariant variables.

An interesting issue which remained unresolved concerns the possibility
of having zero cosmological constant at the present cosmological epoch. In our
model cosmological constant, although in certain sense constrained,
is not automatically zero. Whether or not it can be made equal to zero (exactly
or approximately) without unnatural fine tuning of the parameters is
an open question. Other topics to be elaborated in frames of the theory
considered here are: the origin of
primordial energy density fluctuations, their magnitude and
spectrum, and the universe reheating after inflation. We hope to turn
to these topics in future.

\section*{Acknowledgments}

One of us (Yu.Sh.) would like to express gratitude to Brown University
for hospitality, and especially to thank Robert Brandenberger for
invitation to visit Brown University.

\section*{Appendix}

Our theory is based on Riemann--Cartan space $U_4$  (for  a  good detailed description
see e.g. [29]). This space naturally  arises within the framework of the Poincar\'e
gauge  theory  of  gravity. Riemann-Cartan structure in $U_4$  implies the presence of
the affine connection form  and  the  metric  tensor  which  is  covariantly constant.
Let $\{e_a (x)\}$ be an  arbitrary  field  of  bases  in  the tangent space of $U_4$  at
each point $x$, and $\{e^a (x)\}$ - the  field  of their dual bases, that is
$$
     \left\langle e^a(x),\, e_b(x) \right\rangle \, =  \delta^a_b \, .  \eqno    (A1)
$$
Latin indices run from $0$ to $3$. The affine connection form
$\omega^a{}_b (x)$ referred to
the basis $\{e^a(x)\}$ defines covariant derivative (denoted $\nabla^\omega$)
of tensors in $U_4.$

The metric tensor $g$ can be developed as follows
$$
g  =  g_{ab}e^a \otimes e^b = g_{\mu\nu}dx^\mu \otimes dx^\nu
\,  , \eqno               (A2)
$$
where $g_{ab}(x)$ and $g_{\mu\nu}(x)$ are  symmetric  components. In  Riemann--Cartan
space $U_4$  the affine  connection  form $\omega^a{}_b$    satisfies  the metricity
condition
$$
\nabla^\omega g  =  0 \, . \eqno           (A3)
$$

The metric components $g_{ab}$ ($g_{\mu\nu}$), and their inverse
(in  the sense of matrices)
$g^{ab}$ ($g^{\mu\nu}$) allow one to raise and  lower  Latin
(Greek) indices. The metricity condition (A3) written in terms of
the components $g^{ab}$   of the metric reads
$$
\omega^{(ab)} \equiv {1\over 2} (\omega^{ab} + \omega^{ba} )
=  {1\over 2} dg^{ab} \, . \eqno                   (A4)
$$

Torsion tensor components $Q^a{}_{bc} = Q^a{}_{[bc]} \equiv {1\over 2}
(Q^a{}_{bc} - Q^a{}_{cb})$ are defined by the following relations
$$
de^a + \omega^a{}_b \wedge e^b = - Q^a{}_{bc}e^b \wedge e^c \, .
\eqno (A5)
$$

{}From Eq.~(A3) it follows that
$$
\omega ^a{}_{bc} = \Gamma^a{}_{bc} + Q^a{}_{bc} + Q_{bc}{}^a +
Q_{cb}{}^a \, ,  \eqno                  (A6)
$$
where $\Gamma^a{}_{bc}$ are the components of the Riemannian
connection  form
(which is constructed in the usual way from  the  metric  tensor)
written in the basis $\{e_a\},\, \{e^a\}.$

In  order  to  describe  spinors  one  has  to   choose   an
orthonormal vector basis (tetrad) $\{e_a\}$, in which
$$
g_{ab} = g(e_a,\, e_b) = \eta_{ab} =  \mbox{diag} (-1,\, 1,\, 1,\, 1)
\, . \eqno      (A7)
$$
In this basis the metricity condition (A4) becomes
$$
\omega^{(ab)} =  0,\,\,\, {\rm or}\,\,\,  \omega^{ab} =  \omega^{[ab]}
\, . \eqno                     (A8)
$$
Henceforth  all  the  formulas  will  refer   to   an   arbitrary
orthonormal basis.

The spin connection form which determines parallel transport
and covariant derivative of four-spinors is defined by
$$
\omega = - {1\over 8} \omega^{ab} \, [\gamma_a,\,\gamma_b]
\, , \eqno       (A9)
$$
where $\gamma^a$  are the usual constant Dirac
$\gamma$-matrices, which satisfy
$$
\{\gamma^a,\,\gamma^b\} = - 2\eta^{ab} \, . \eqno       (A10)
$$
The covariant derivative of a spinor is then defined as follows
$$
\nabla^\omega\psi = d\psi + \omega\psi \, . \eqno     (A11)
$$
It transforms like spinor under the action of local Lorentz group
$L_6.$

For any group  $G$  of intrinsic gauge transformations by  $A =
A_\mu dx^\mu$  we will denote its connection  form.  Then  both  locally
Lorentz and $G$-invariant derivative of any  field  multiplet  $f(x)$
will be
$$
Df = \nabla^\omega f + Af \, , \eqno                    (A12)
$$
where $A$ implies the matrix of the corresponding representation of
the algebra of the group $G.$

Let  us  consider  now  the   group   of   local   conformal
transformations whose action on the metric $g$ is
$$
g(x) \rightarrow  g^\prime (x) = \exp (2\sigma(x))\, g(x)
\, , \eqno                     (A13)
$$
where $\sigma(x)$ is an arbitrary smooth function. Eq.~(A13)  means  that the metric $g$
has conformal weight two:\footnote{A field $f(x)$ will be said to have conformal weight
$w(f)$ if under the action of local conformal group it transforms as
$$
f(x) \rightarrow f^\prime (x) = \exp \left( w(f)\sigma (x) \right) f(x).
$$ }
$$
w(g)  =  2 \, . \eqno                                 (A14)
$$
{}From (A1) and (A7) we immediately obtain conformal weights of the
tetrad basis vectors and their dual one-forms
$$
w(e_a) = - 1, \,\,  w(e^a) = 1 \, .  \eqno                 (A15)
$$

For the affine and gauge connection forms we are to set
$$
w(\omega^a{}_b) = 0,\,\,   w(A)  =  0 \, , \eqno        (A16)
$$
and conformal weights of scalar and spinor fields are, as usual,
$$
w(\varphi) = - 1, \,\,  w(\psi) = - {3\over 2} \, .  \eqno (A17)
$$

Given the affine connection form $\omega^a{}_b$ we  can  construct  its Riemann--Cartan
curvature two-form
$$
R^a{}_b (\omega) = d\omega^a{}_b + \omega^a{}_c \wedge
\omega^c{}_b \, , \eqno                         (A18)
$$
which is conformally invariant and whose components  relative  to
the coordinate basis $\{dx^\mu\}$
$$
R^a{}_{b\mu\nu} (\omega) = \partial_\mu \omega^a{}_{b\nu} - \partial_\nu
\omega^a{}_{b\mu} + \omega^a{}_{c\mu}\omega^c{}_{b\nu} -
\omega^a{}_{c\nu}\omega^c{}_{b\mu} \eqno  (A19)
$$
constitute  those  of  curvature  tensor.  From  this  last   one
constructs the scalar curvature
$$
R(\omega) = R^{ab}{}_{\mu\nu}(\omega)e^\mu_a e^\nu_b \, , \eqno (A20)
$$
whose conformal weight is
$$
w(R(\omega)) = - 2 \, . \eqno                            (A21)
$$

{}From Eq.~(A5) there follows the transformation  law  for  the torsion tensor components
$$
Q^a{}_{bc}(x) \rightarrow Q^{\prime a}{}_{bc}(x) =
\exp (- \sigma (x)) \left( Q^a{}_{bc}(x) + \delta^a{}_{[b}
<d\sigma (x),\, e_{c]} (x)> \right) \, . \eqno  (A22)
$$
As it can be shown these transformations affect only  the  vector
trace part of the torsion tensor
$$
Q_a = Q^b{}_{ab} =  Q_\mu e^\mu_a \,  ,  \eqno       (A23)
$$
so that
$$
Q_\mu (x) \rightarrow  Q^\prime_\mu (x) = Q_\mu (x) - {3\over 2}
\partial_\mu \sigma (x) \, . \eqno               (A24)
$$

Locally Lorentz and $G$-invariant derivative $Df$ defined by the
expression (A12) does not preserve conformal  properties  of  the
fields. Using transformation  law  (A24)  of  the  torsion  trace
vector and taking into account (A16) and (A17) we can construct a
new derivative
$$
{\cal D} f  =  Df + {2\over 3} w(f)Qf  \eqno               (A25)
$$
with the desired property
$$
w({\cal D} f) = w(f) \, ,  \eqno             (A26)
$$
where $Q$ is the one-form of the torsion trace
$$
Q  =  Q_a e^a =  Q_\mu dx^\mu \, .  \eqno            (A27)
$$

The derivative ${\cal D}$  is  used  in  our  paper  in  constructing
locally conformally invariant actions.


\section*{References}

\begin{thebibliography}{99}
\bibitem{one}
Weyl H. (1918) {\it Sitzung. d. Preuss. Akad. d. Wiss.} 465; \\
     Weyl H. (1920) {\it Raum. Zeit. Matterie.} (Berlin: Springer).
\bibitem{two}
Dirac P.A.M. (1973) {\it Proc. Roy. Soc. London} {\bf A333}, 403.
\bibitem{three}
Utiyama R. (1973) {\it Prog. Theor. Phys.} {\bf 50}, 2080.
\bibitem{four}
Freund P.G.O. (1974) {\it Ann. Phys.} (N.Y.) {\bf 84}, 440.
\bibitem{five}
Utiyama R. (1975) {\it Prog. Theor. Phys.} {\bf 53}, 565.
\bibitem{six}
Hayashi K., Kasuya M. and Shirafugi T. (1977) {\it Prog. Theor. Phys.} {\bf 57}, 431.
\bibitem{seven}
Hayashi K. and Kugo T. (1979) {\it Prog. Theor. Phys.} {\bf 61}, 334.
\bibitem{eight}
Cheng H. (1988) {\it Phys. Rev. Lett.} {\bf 61}, 2182.
\bibitem{nine}
Wheeler J.T. (1990) {\it Phys. Rev.} {\bf D41}, 431.
\bibitem{ten}
Hochberg D. and Plunien G. (1991) {\it Phys. Rev.} {\bf D43}, 3358.
\bibitem{eleven}
Zhytnikov V.V. (1993) {\it Int. J. Mod. Phys.} {\bf A8}, 5141.
\bibitem{twelve}
Nieh H.T. and Yan M.L. (1982) {\it Ann. Phys.} (N.Y.) {\bf 138}, 237.
\bibitem{thirteen}
Obukhov Yu.N. (1982) {\it Phys. Lett.} {\bf A90}, 13; \\
     Dereli T. and Tucker R.W. (1982) {\it Phys. Lett.} {\bf B110}, 206.
\bibitem{fourteen}
Gibbons G.W. and Hawking S.W. (1977) {\it Phys. Rev.} {\bf D15}, 2752.
\bibitem{fifteen}
Linde A.D. (1990) {\it Particle physics and inflationary cosmology}. (Chur, Switzerland:
Harwood Academic Publishers).
\bibitem{sixteen}
Misner C.N., Thorn K.S. and Wheeler J.A. (1973) {\it Gravitation}. (San Francisco:
Freeman).
\bibitem{seventeen}
Alvarez E. (1989) {\it Rev. Mod. Phys.} {\bf 61}, 561.
\bibitem{eighteen}
Halliwell J.J. (1990) {\it Introductory lectures on quantum cosmology}, in: {\it
Proceedings of the Jerusalem Winter School on Quantum Cosmology and Baby Universes} (Ed.
by T.Piran).
\bibitem{nineteen}
Duncan M.J. (1990) {\it Quantum geometrodynamics}. Lectures presented at TASI summer
school; UMN-TH-916/90.
\bibitem{twenty}
Vilenkin A. (1984) {\it Phys. Rev.} {\bf D30}, 509.
\bibitem{twentyone}
Vilenkin A. (1986) {\it Phys. Rev.} {\bf D33}, 3560.
\bibitem{twentytwo}
Vilenkin A. (1988) {\it Phys. Rev.} {\bf D37}, 888.
\bibitem{twentythree}
Hawking S. (1982) in {\it Astrophysical cosmology: Proceedings of the study week on
cosmology and fundamental physics}, Eds. H.A.Bruck, G.V.Coyne and M.S.Longair (Vatican:
Pontificae Academiae Scientarium Scripta Varia).
\bibitem{twentyfour}
Hawking S. (1983) in {\it Relativity, Groups and Topology II}, Eds.
     B.S.DeWitt and R.Stora (Amsterdam: North Holland Physics
     Publishing).
\bibitem{twentyfive}
Hartle J.B. and Hawking S.W. (1983) {\it Phys. Rev.} {\bf D28}, 2960.
\bibitem{twentysix}
Hawking S. (1984) {\it Nucl. Phys.} {\bf B239}, 257.
\bibitem{twentyseven}
Halliwell J.J. and Louko J. (1989) {\it Phys. Rev.} {\bf D39}, 2206.
\bibitem{twentyeight}
Vilenkin A. (1989) {\it Phys. Rev.} {\bf D39}, 1116.
\bibitem{twentynine}
Hehl F.W., Von der Heyde P., Kerlick G.D. and Nester J.M. (1976) {\it Rev. Mod. Phys.}
{\bf 48}, 393.
\end{thebibliography}
\end{document}